 \documentclass[preprint2]{aastex}

\slugcomment{submitted to Ap. J. Letters}

\shorttitle{Limit on the Age of the Universe}
\shortauthors{L.M. Krauss}

\begin{document}


\title{Dark Energy and the Hubble Age}


\author{L.M. Krauss}
\affil{Departments of Physics and Astronomy, Case Western Reserve University, 
Cleveland OH 44106-7079}
\email{krauss@cwru.edu}



\begin{abstract}
I point out that an effective upper limit of approximately 20 Gyr (for a Hubble constant 
of 72 km/s/Mpc) or alternatively on the $H_0$-independent quantity $H_0t_0 < 1.47$, exists on the age of the Universe, essentially independent of the unknown equation of state of the dominant dark energy
component in the Universe.   Unless astrophysical constraints on the
age of the Universe can convincingly
reduce the upper limit  to below this value no 
useful lower limit on the equation of state parameter $w$ for this component can be
obtained.  Direct dating by stars does not provide a useful constraint, but model-dependent cosmological limits from supernovae and the CMB observations may.  For a constant value of $w$, a bound $H_0t_0 < 1.1$ gives a limit $ w> -1.5$

\end{abstract}




\section{Introduction}

The realization that some unknown form of energy density associated with otherwise empty space appears to dominate the gravitational dynamics of the Universe has changed virtually everything in cosmology.   For example, the future evolution of the Universe becomes largely independent of its geometry (\cite{kraussturn}), so that a closed universe can expand forever, and an open universe can ultimately collapse.   

One of the earliest motivations for assuming the existence a cosmological constant,  the simplest form of dark energy, involved a comparison of the Hubble age of the Universe---determined for an assumed cosmological model on the basis of the observed expansion rate today---with a lower limit on the age of the oldest objects in our galaxy.   In order to resolve the paradox when the latter exceeded the former (\cite{demarque}), cosmologists were driven to consider the possibility of exotic cosmological models that might allow an older universe for a fixed value of the Hubble constant today.  

An accelerating universe allows for this possibility simply because galaxies that are now located at a certain distance from us, and which are moving at some fixed velocity were separating from us at a smaller velocity at earlier times, and thus would have required longer to achieve their present separation than would otherwise be required.   For a flat Universe, in the limit where a cosmological constant dominates the energy density, the Hubble age can approach infinity for any value of the Hubble constant.

We currently have no fundamental understanding of the nature of the dark energy (or perhaps more accurately dark "pressure").  It is significant that a lower limit on the age of the Universe determined from globular cluster dating techniques now provides independent evidence for the  existence of dark energy, and puts a limit on the equation of state parameter $wp/\rho$ ( i.e. w=pressure/energy density) of the dark energy $w <-0.4$ (\cite{krausschab}).  The question also arises however, if exotic equations of state for dark energy can increase the Hubble age, can one put useful constraints of such equations of state from an {\it upper} limit on the age of globular clusters, or from direct estimates of the Hubble age itself.  I investigate these questions here.

\section{Exotic Equations of State and The Hubble Age in a Flat Universe }

Determination of the distance-redshift relation made using distant Type 1a supernovae (\cite{perl,kirsh}), combined with independent estimates for both the mass density
in the Universe today, and the geometry of the universe from CMB measurements (\cite{boomerang,maxima}) have definitively established the need for a dominant component to the energy budget that involves a negative pressure.  

While there is currently no fundamental understanding of the nature of this dark pressure, one particular value of the equation of state parameter carries special weight.  A vacuum energy density is fixed, by Lorentz invariance, to have the form $T_{\mu \nu} = \Lambda g_{\mu \nu}$ and thus as $w=-1$.    Unfortunately, however, all estimates of the vacuum energy density on the basis of first principles calculations yield a value which is orders of magnitude too large, and thus it was commonly  assumed for many years that some new symmetry mechanism might yield a value for the vacuum energy which is precisely zero.    A uniform scalar field, for example, that is slowly rolling down a potential and has not yet achieved its minimum value can mimic vacuum energy.  For such a field, w is given by

\begin{equation}
\nonumber
w={{\dot\phi^2/2 -V} \over {\dot\phi^2/2 +V } }
\end{equation}

Since the kinetic energy of the scalar field as it rolls in the potential gives a positive contribution to the pressure, any rolling implies $w > -1$.   

Of course, since we do not have any underlying theory for the dark pressure one must allow for  the possibility that $w <-1$ (\cite{caldwell}).  It is clear that Lagrangian models that have an equation of state of this form will be extremely exotic, implying for example, a negative kinetic term.  Such models will have the remarkably odd feature that the energy density of the dark energy will {\it increase} with time!  As a result, the Hubble constant itself will continue to increase with time.

If a cosmological constant allows for an older universe for a fixed Hubble constant today, what will be the effect of even more exotic forms of dark pressure?    If the equation of state parameter remains constant, for  a fixed universe, the age-Hubble constant relation is given by:

\begin{eqnarray}
 { \nonumber H_0t_0 =  \hspace{60mm} } \\ \nonumber \\
\nonumber \int _{0}^{\infty }{dz \over{(1+z)} 
[(\Omega_{m})(1+z)^3  +  (\Omega_X)(1+z)^{3(1+w)}]^{1/2}}
\end{eqnarray}

\noindent{where $\Omega_{m}$ is the fraction of the closure density in matter today, and $\Omega_X$ is the fraction of the closure density in material with an equation of state parameter $w$.}

It is clear that as $\Omega_X$ approaches unity, the age of the Universe can approach infinity if $w \le -1$.   However, we have good estimates on the density of dark matter today, coming from gravitational lensing of clusters \cite{tony}, X-Ray studies of clusters\cite{evrard}, and studies of large scale structure \cite{sloan,2df}, that conservatively imply $\Omega_m \ge 0.2$.    If we assume this minimal value, then the implications of the above relation between age and Hubble constant for exotic forms of energy become quite different.   

If we normalize to the Hubble Key Project best fit value  $H_0 =72$ km s$^{-1}$ Mpc$^{-1}$ (\cite{keyproj}) then we can plot the above relation for age as function of $w$, shown in figure 1.    Also shown in this figure is the Hubble-independent product $H_0t_0$.  As is clearly seen in the figure the age of the Universe is a sharply increasing function of $-w$ for $w < 0$, but then it quickly begins to asymptote, so that for $w < -10$ the age increases by less than 0.5 Gyr for $w>-30$!

This behavior is easily understood.   As has been descibed, as $w$ decreases below -1, the net energy density stored increases with time.   Thus, the relative contribution of this exotic energy to the total energy budget of the Universe was smaller at earlier times (higher redshifts) than, say, the energy density stored in a cosmological constant.  In short, this exotic energy has 'just' become important.  As a result, the more negative is $w$, the less time there has been for it to have an effect, even though the acceleration rate increases during the period in which it is significant.  The net result is that, for fixed fraction of the closure density today in matter, there is effectively a maximum age for the universe, independent of how negative $w$ is!  For $H_0=72$ today, one finds, for example, that
for $w >-600$, $t_0 <20 $Gyr.  Put in $H_0$-independent terms, one finds $H_0t_0 \le 1.47$. 

( One can also derive an asymptotic upper bound $H_0t_0 < {2 / (3 \Omega_M^{1/2})} (=1.49$ for $\Omega_M=0.2$) (valid also for non-constant $w$) by taking the limit of the equation above for $H_0t_0$ as $w$ approaches negative infinity\footnote{This result was independently suggested for inclusion here by Chiba (for constant
w) and Repko (for varying w) (\cite{chibarepo}) following the posting of the original astro-ph version of this article.}.)

\section{Limiting $w$ From Age Considerations }

From the above analysis, it is clear that in order to derive a robust constraint on negative values of the equation of state parameter $w$, one must be able to place an upper limit on the age of the universe in the range of 15-18 Gyr, $H_0=72$.   The sharp rise of age with negative $w$ tails off considerably above 18 Gyr.

Unfortunately, at the current time such a direct determination of the upper limit on the age of stellar systems is not possible.   Recent Monte Carlo studies of stellar age constraints in old globular clusters yield a $95 \%$ upper limit of $16$ Gyr (\cite{krausschab}).  If this were the end of the story, then some useful constraint would be derivable.  However, to this upper limit one must add a conservative upper limit on the time between the Big Bang and the formation of stars in our galaxy.  While the epoch of first star formation is likely to be at $z >6$, galaxy formation continues down to redshifts as low as 1-2.   This implies that the first stars in our galaxy could have started forming when the Universe was as old as 4-5 Gyr.  This not likely, but it is {\it possible}.    While  new techniques (i.e. see \cite{krausschab2}) may allow the upper limit on globular cluster ages to decrease, it is difficult to imagine ways to reduce this latter 5 Gyr uncertainty on the period before the formation of our galaxy.  

As a result, the most robust age constraints on $w$ (or on the time-weighted integral of $w$, if $w$ varies over cosmic time) will probably come from direct lower limits on $t_0$ itself, but from cosmological estimates of $t_0$ or the combination $H_0t_0$,  which can be directly probed by redshift vs distance measures (i.e. \cite{reiss}) and CMB experiments, for example (i.e. \cite{cmbconst}).   These depend to some extent on cosmological parameter estimation, and on uncertainty in cosmological models (which become more uncertain if $w$ is not constant), but early estimates already suggest limits of $H_0t_0 <1.1$ may  be possible.  This would put a constraint on constant $ w >-1.5$.

Equations of state with $w <-1$ violate the weak equivalence principle, and thus have not been examined theoretically in great detail.  One might hope therefore, that observational constraints could provide some significant guidance for theorists in this regard.  Unfortunately, direct age determinations have a residual uncertainty which is unlikely to allow significant constraints to be derived.  Instead,  cosmological estimates using supernovae and CMB data appear to offer the best possibility for constraining large negative values of $w$.

At the same time, it is significant that cosmology implies an upper limit on the age of the Universe that is essentially independent of the unknown value of $w$.  This allows a globular cluster ages, at the very least, to provide a direct consistency test of our fundamental cosmological framework.  While unlikely, a direct age determination in excess of $20$ Gyr would be inconsistent with the Hubble Age for any cosmic equation of state, for a Hubble constant of $72$..

The author acknowledges support from the DOE and NSF, and useful conversations with Robert Caldwell. This work was initiated at the KITP at Santa Barbara. 





\begin{figure}
\plotone{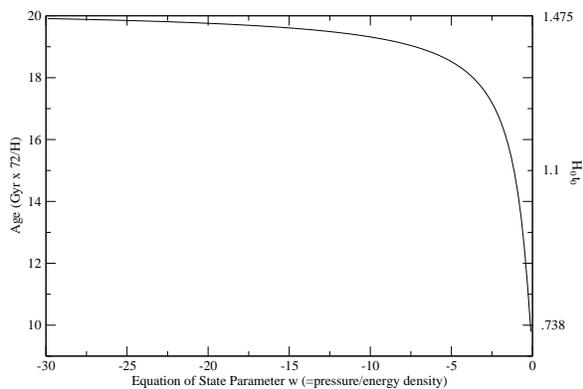}
\caption{Age of the universe as a function of the equation of state parameter, $w$, for constant equation of state. \label{fig1}}
\end{figure}



\begin{thebibliography}{}
\bibitem[Caldwell (2002)]{caldwell} Caldwell, R., 2002, Phys. Lett. B545, 23
\bibitem[Chaboyer and Krauss (2002)]{krausschab2} Chaboyer, B., and Krauss, L.M., 2002, \apj, 567, L45
\bibitem[de Bernardis et al (2000)]{boomerang}  de Bernardis, P. {\it et al}, 2000, Nature, 404, 995
(2000)
\bibitem[Freedman et al (2001)]{keyproj} Freedman, W. {\it et al}, \apj, 553, 47
\bibitem[Hanany et al (2000)] {maxima}Hanany, S. {\it et al}, 2000, \apj,  545, 5L 
\bibitem[Janes and Demarque (1983)] {demarque} Janes, K., Demarque. P.,  1983, \apj,
    264, 206
\bibitem[Knox et al (2001)]{cmbconst} Knox, A., {\it et al}, 2001,  \apj, 563, L95
\bibitem[Krauss and Chaboyer (2003)] {krausschab} Krauss, L.M., and Chaboyer, B,  2003, Science, in press
\bibitem[Krauss and Turner (1999)] {kraussturn} Krauss, L.M., and Turner, M. S..,  1999, J. Gen. Rel. Grav, 31, 1543
\bibitem[Perlmutter et al (1999)]{perl} Perlmutter, S., {\it et al},1999, \apj, 517, 565 
\bibitem[Riess et al (1998)]{reiss} Riess, A., {\it et al}, 1998, \aj, 116, 1009
\bibitem[Schmidt  et al (1998)]{kirsh} Schmidt, B., {\it et al}, 1998, \apj, 507, 46
\bibitem[T. Chiba; W. Repko (2003)]{chibarepo}T. Chiba, Private communication; W. Repko, Private Communication. 
\end{thebibliography}
\end{document}